\documentclass{article}

\usepackage{arxiv}

\usepackage[utf8]{inputenc} 
\usepackage[T1]{fontenc}    
\usepackage{url}            
\usepackage{booktabs}       
\usepackage{amsfonts}       
\usepackage{nicefrac}       
\usepackage{microtype}      
\usepackage{lipsum}		
\usepackage{graphicx}
\usepackage[square,numbers]{natbib}
\usepackage{doi}
\usepackage{amsmath}

\usepackage{algorithm, algpseudocode}
\usepackage{svg}

\title{Interpolating Rotations with Non-abelian Kuramoto Model on the 3-Sphere}

\author{  Zinaid Kapić, Aladin Crnkić \\
	Faculty of Technical Engineering \\
	University of Bihać \\
    Bihać \\
	\texttt{zinaid.kapic@unbi.ba, aladin.crnkic@unbi.ba}
}

\renewcommand{\shorttitle}{Interpolating Rotations with Non-abelian Kuramoto Model on the 3-Sphere}

\hypersetup{
pdftitle={Interpolating Rotations with Non-abelian Kuramoto Model on the 3-Sphere},
pdfsubject={Computer Science},
pdfauthor={Zinaid Kapic, Aladin Crnkic},
pdfkeywords={Rotations, Quaternion, Interpolation, Non-Abelian Kuramoto model},
}

\begin{document}
\maketitle

\begin{abstract}
	The paper presents a novel method for interpolating rotations based on the non-Abelian Kuramoto model on sphere \({\mathbb{S}}^{3}\). The algorithm, introduced in this paper, finds the shortest and most direct path between two rotations. We have discovered that it gives approximately the same results as a Spherical Linear Interpolation algorithm. Simulation results of our algorithm are visualized on \({\mathbb{S}}^{2}\) using Hopf fibration. In addition, in order to gain a better insight, we have provided one short video illustrating the rotation of an object between two positions.
\end{abstract}


\section{Introduction}
Data in the form of 3-dimensional rotations are applied in many spheres of computer science, kinematics, robotics, computer vision, etc. The most common problems with 3-dimensional rotation data are its averaging and interpolation. The problem of rotation averaging has been extensively studied in the last decade by many authors \cite{Hartley2013, Kapic2021}.

In this paper, we have studied rotation interpolation problem which is one of the fundamental problems in computer animation, robotics and structure from motion \cite{Kang1999,Jttler1994, Park1997}. Due to the need for smooth interpolation between camera frames, it is often used in computer graphics \cite{Haarbach2018}. It also finds application in animations where each joint of a body is usually represented by a rotation so animations in keyframe are created interpolating between these rotations \cite{Dam2000QuaternionsIA}. Furthermore, it is worth mentioning that motion control is required in robotics, where interpolation is used for obstacle avoidance and object approach, such as 'pick and place' examples of robot movement \cite{Svejda2015}. 

There are several representations of the rotations such as rotational matrices, Euler angles, and quaternions \cite{Hartley2013}. Matrix representation of rotations is not a recommended way for interpolation purposes because simple linear combinations of the coefficients can sometimes result in non-positive orthogonal matrices, and therefore non-valid rotations. Euler angles are better than rotation matrices, but the main disadvantage of this representation is a Gimbal lock problem \cite{Dam2000QuaternionsIA}. 

Quaternions arise as the most efficient way of rotation representation, especially for rotation interpolation. Quaternions represent a number system that expands complex numbers and they are defined as \cite{Gnati2012QuaternionsAT}:
\begin{equation*}
q=a+bi+cj+dk,
\end{equation*}
where $a,b,c$ and $d$ are real numbers and $i^2=j^2=k^2=ijk=-1$. A special group of quaternions is unit quaternions that correspond to a rotation and represent a point on unit sphere \({\mathbb{S}}^{3}\) in 4-dimensional space. In other words, if the Euclidean norm
\begin{equation*}
||q||=\sqrt{a^2+b^2+c^2+d^2}
\end{equation*} 
is equal to 1, then a quaternion is called a unit quaternion \cite{Gnati2012QuaternionsAT}.

Quaternion inverse is defined as $q^{-1}=\bar{q}/||q||^2$, where $\bar{q}=a-bi-cj-dk$ is a conjugated quaternion. The equation $q^{-1}q=qq^{-1}=1$ further defines that $q^{-1}=\bar{q}$ is valid for each unit quaternion. Quaternion popularity rose with aircraft simulations and computer graphics where quaternions were used for describing 3-dimensional rotations \cite{Gnati2012QuaternionsAT}.

Throughout this paper we will limit ourselves on interpolation between two rotations. The most common and a standard way of rotation interpolation is a Spherical Linear Interpolation (SLERP) algorithm [10].  Using this method, interpolation from unit quaternion $p$ to unit quaternion $r$ with parameter $t \in [0,1]$ is defined as \cite{Shoemake1985}:

\begin{equation*}
Slerp(p,r;t)=p(p^{-1}r)^t.
\end{equation*}

A novel method for rotation interpolation has been proposed in this paper. This
method is based on generalizations of the famous Kuramoto model to higher dimensions. Kuramoto model represents the most significant model for studying collective behavior and self-organization in large populations of coupled oscillators \cite{Kuramoto}. One of the generalizations of this model has been introduced by Lohe and is known as the non-Abelian Kuramoto model \cite{Lohe2009}. Authors in the paper \cite{Jaimovi2018} used a variation of this model in quaternion form for studying low-dimensional dynamics. That model is defined as

\begin{equation}
q_i=q_ifq_i+wq_i+q_i u -\bar{f}, i=\bar{1,N},
\end{equation}

where $f=f(t,q_1, \dots q_N)$ is a quaternionic function that represents the coupling between particles \cite{Jaimovi2018}. This model is a non-Abelian Kuramoto model on the 3-sphere. 

Different choices of coupling functions in the model (1) solve various problems in different scientific fields. For instance, this model is useful for coordination and consensus on groups \({\mathbb{S}}^{3}\) and SO(3) \cite{Crnki2020}, clustering of static and stream data \cite{Crnki2018, Crnki2020Swarms}, and other applications in science and engineering \cite{ha2020dynamical}.

The paper is organized as follows. In the next section, we will explain in more detail our algorithm for interpolating rotations. In Section 3 we present simulation results that illustrate our method and visualize them as curves on \({\mathbb{S}}^{2}\) using Hopf fibration. This section will also provide a video for visualization of object’s rotation interpolation. Finally, in the last section we will present our conclusion, possible upgrades of this method, and further research.
\section{Algorithm}
In this section we will introduce our method for interpolating rotations using nonAbelian Kuramoto model on \({\mathbb{S}}^{3}\). Let us primarily suppose that the rotations $p$ and $r$ are represented by unit quaternions. Setting $N= 1$ and $u= w= 0$ in (1), we get following quaternion-valued ordinary differential equation
\begin{equation}
q^{'}(t)=q(t)fq(t)-\bar{f},
\end{equation}
where $f$ is a quaternionic coupling function. Note that differential equation (2) preserves the unit sphere \({\mathbb{S}}^{3}\). This means that if the initial condition of equation (2) satisfies $q(0) \in \mathbb{S}^{3}$ then the solution to that equation satisfies $q(t) \in \mathbb{S}^{3}$ at any moment $t$. We can observe the last equation (2) as matrix ordinary differential equation on $SU(2)$ because the group of unit quaternions is isomorphic to the group of special
unitary matrices $SU(2)$.

By substituting $f=-\frac{1}{2}(q(t)\bar{r}q(t)-r)$, in (2) and using initial condition $q(0)=p$, one obtains
\begin{equation}
q^{'}(t)=-\frac{1}{2}(q(t)\bar{r}q(t)-r),
\end{equation}
which is a dynamical model that will be used in our algorithm. Solution to this differential equation is an interpolating smooth curve $q(t) \in \mathbb{S}^{3}$ which connects the initial unit quaternion $p$ and the final one $r$ In other words, the initial condition of model (3) is given by the initial unit quaternion $(q(0)=p$, and as time $t$ increases, the solution curve $q(t)$ tends to the final unit quaternion $(r=\lim\limits_{t \to \infty} q(t))$.

If we substitute $q(t)=q_1(t)+q_2(t)i+q_3(t)j+q_4(t)k$, $p=p_1+p_2i+p_3j+p_4k$ and $r=r_1+r_2i+r_3j+r_4k$ in (4), we will obtain the following system of ordinary differential equations:
\begin{equation}
\Bigg\{
 \begin{matrix}
  q_1^{'}=r_1-q_1(t)-q_1(t)cos\theta(t) \\
   q_2^{'}=r_2-q_2(t)-q_2(t)cos\theta(t) \\
   q_3^{'}=r_3-q_3(t)-q_3(t)cos\theta(t) \\
    q_4^{'}=r_4-q_4(t)-q_4(t)cos\theta(t)
 \end{matrix}
\end{equation}

with initial conditions $q_1(0)=p_1, \dots, q_4(0)=p_4$. This model is more convenient for simulating than (3). In system (4), $cos\theta(t)$ is an inner product (dot-product) of unit quaternions $q(t)$ and $r$, or mathematically expressed as $cos\theta(t)=q(t) \cdot r = q_1(t)r_1+q_2(t)r_2+q_3(t)r_3+q_4(t)r_4$. 

Furthermore, we will explain our method for interpolating between two rotations in more detail through the algorithm. We tend to refer to this algorithm as KLI (Kuramoto-Lohe Interpolation) algorithm.

The KLI algorithm is as follows:
\begin{algorithm}
\caption{KLI algorithm for interpolating between two rotations}
\begin{algorithmic}[1]
\State Enter $p$, $r$
\State Choose tolerance $\varepsilon$, step $\delta$, and define $T=0$
\State Solve (3) with $q(0)=p$
\Loop
  \If{||$r-q(T)||<\varepsilon$}
    \State \textbf{return} $q(t)$, for $t \in [0,T]$
  \Else
    \State $T=T+\delta$
  \EndIf
\EndLoop
\end{algorithmic}
\end{algorithm}
 
\section{Simulations}
This section presents simulation results of our method. We will assume that the initial rotation (represented by unit quaternions) has value $p=0+0i+0j+1k$ and the final one $r=0.5+0.5i+0.5j+0.5k$. The goal is to find interpolating  curve between $p$ and $r$, which lies on the surface of the unit sphere \({\mathbb{S}}^{3}\) in the 4-dimensional space. Since such curve is problematic to visualize, we will use Hopf’s fibration to describe it as a curve lying on a unit sphere \({\mathbb{S}}^{2}\) in the 3-dimensional space. In 1931, Heinz Hopf defined this mapping in the paper \cite{Hopf1931}, and since then it has found applications in magnetic monopoles, rigid body mechanics, and quantum information theory. Hopf fibration is defined as a mapping $h:\mathbb{S}^{3} \to \mathbb{S}^{2}$ \cite{Lyons2003}:

\begin{equation}
h(a,b,c,d)=(a^2+b^2-c^2-d^2, 2(ad+bc), 2(bd-ac)).
\end{equation}

Using the KLI algorithm with tolerance $\varepsilon = 10^{-5}$ and step $\delta = 0.01$, interpolating curve $q(t)$ reached the final unit quaternion at moment $t = 11.66$, as shown in Figure 1a). From closer inspection of Figure 1a) and Figure 1b), we can see that KLI and SLERP algorithms give approximately the same results. 

\begin{figure}[!htbp]
	\centering
    \includegraphics[width=0.65\columnwidth]{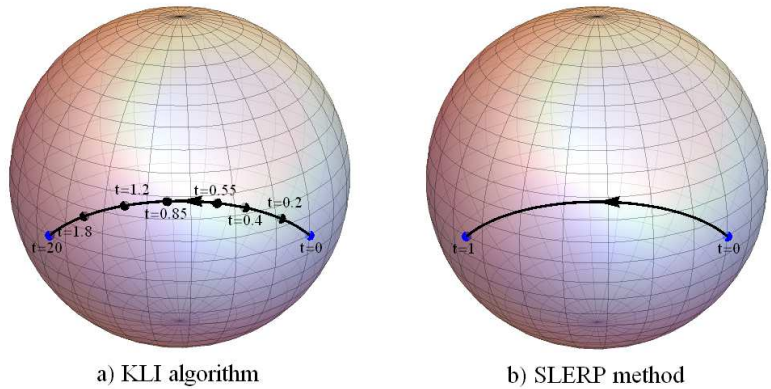}
	\caption{Interpolation curves obtained with KLI and SLERP algorithms.}
	\label{fig:fig1}
\end{figure}

Table 1 presents quaternion values of interpolating curve $q(t)$ at different moments. We have visualized those values as thick points in Figure 1a).

\begin{table}[!htbp]
	\caption{Quaternion values of q(t) at different moments t}
	\centering
	\begin{tabular}{cc}
		\toprule
		\cmidrule(r){1-2}
		\textbf{t}     & \textbf{q(t)} \\
		\midrule
		\\
		0 & $0.00+0.00i+0.00j+1k$       \\ \\
		0.2 & $0.09+0.09i+0.09j+0.98k$       \\ \\
		0.4 & $0.17+0.17i+0.17j+0.95k$       \\ \\
		0.55 & $0.23+0.23i+0.23j+0.92k$       \\ \\
		0.85 & $0.31+0.31i+0.31j+0.84k$       \\ \\
		1.2 & $0.37+0.37i+0.37j+0.76k$       \\ \\
		1.8 & $0.37+0.37i+0.37j+0.76k$       \\ \\
		11.66 & $0.49+0.49i+0.49j+0.5k$       \\ \\
		\bottomrule
	\end{tabular}
	\label{tab:table1}
\end{table}
Figure 2 illustrates the evolution of interpolation by displaying rotating object in 3-dimensional space at the same moments provided in Table 1. For this purpose, we have used polar form of unit quaternion defined as

\begin{equation*}
q=cos(\frac{\theta}{2})+Usin(\frac{\theta}{2})
\end{equation*}

where $U=q/||q||$ and $tg(\theta/2)=||q||/q_0$ \cite{Sangwine2008}. In that way, the object’s rotation can be specified by the unit vector $U$ that defines an axis of rotation and the rotation angle $\theta$ around that axis. However, such representation can cause some peculiar effects, since two antipodal quaternions $q$ and $-q$ correspond to the same rotation.

\begin{figure}[!htbp]
	\centering
    \includegraphics[width=1.0\columnwidth]{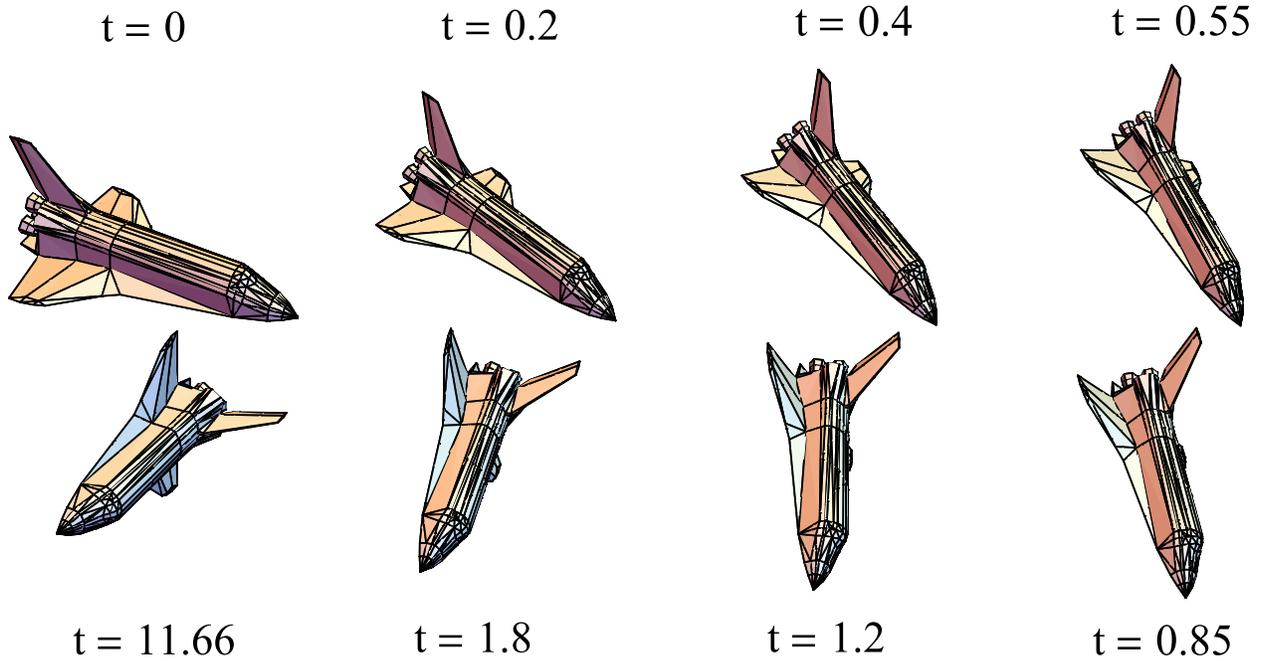}
	\caption{ Interpolation of object’s rotation. }
	\label{fig:fig2}
\end{figure}

Furthermore, for the purpose of better illustration, we have provided one video demonstrating the continuous motion from $p$ to $r$.\footnote{Supplementary video that demonstrates rotation interpolation can be found at \url{https://www.youtube.com/watch?v=IkE_olufAuI}.}
\\
\\
\noindent{\textbf{Acknowledgements}}
The second author acknowledges the support of the COST action CA16228 “European Network for Game Theory”.

\section{Conclusion}
The paper introduces a novel method for performing an interpolation between two
rotations. This method is based on the so-called non-Abelian Kuramoto model. Simulation results of our algorithm are visualized on  \({\mathbb{S}}^{2}\) using Hopf fibration and illustrated on a rotating object. Furthermore, to gain better insight, we have provided one short video demonstrating the continuous motion of that object.

The simulation results indicate that our algorithm gives approximately the same results as the SLERP algorithm finding the shortest and most direct path between two rotations. The algorithm proposed in this paper can be used, for instance, for various cases in robotics, robotic hand movement, motion control and object avoidance. For that purpose, it would be interesting to implement this algorithm in ROS (Robot Operating System). ROS represents a meta-operating system for writing robot software that offers users a way to quickly build, maintain and expand their robot \cite{Estefo2019}. In addition, it can find application in animations in computer graphics.

It is noticeable that this algorithm can be used only for interpolation between two rotations. Hence, we find it interesting to introduce a model that will interpolate multiple rotations. We will address this open question in further research.

\bibliographystyle{unsrtnat}
 


\begin{thebibliography}{21}
\providecommand{\natexlab}[1]{#1}
\providecommand{\url}[1]{\texttt{#1}}
\expandafter\ifx\csname urlstyle\endcsname\relax
  \providecommand{\doi}[1]{doi: #1}\else
  \providecommand{\doi}{doi: \begingroup \urlstyle{rm}\Url}\fi

\bibitem[Hartley et~al.(2013)Hartley, Trumpf, Dai, and Li]{Hartley2013}
Richard Hartley, Jochen Trumpf, Yuchao Dai, and Hongdong Li.
\newblock Rotation averaging.
\newblock \emph{International Journal of Computer Vision}, 103\penalty0
  (3):\penalty0 267--305, January 2013.
\newblock \doi{10.1007/s11263-012-0601-0}.
\newblock URL \url{https://doi.org/10.1007/s11263-012-0601-0}.

\bibitem[Kapic et~al.(2021)Kapic, Crnkic, Jacimovic, and Mijajlovic]{Kapic2021}
Zinaid Kapic, Aladin Crnkic, Vladimir Jacimovic, and Nevena Mijajlovic.
\newblock A new dynamical model for solving rotation averaging problem.
\newblock In \emph{2021 20th International Symposium {INFOTEH}-{JAHORINA}
  ({INFOTEH})}. {IEEE}, March 2021.
\newblock \doi{10.1109/infoteh51037.2021.9400663}.
\newblock URL \url{https://doi.org/10.1109/infoteh51037.2021.9400663}.

\bibitem[Kang and Park(1999)]{Kang1999}
I.~G. Kang and F.~C. Park.
\newblock Cubic spline algorithms for orientation interpolation.
\newblock \emph{International Journal for Numerical Methods in Engineering},
  46\penalty0 (1):\penalty0 45--64, September 1999.
\newblock
  \doi{10.1002/(sici)1097-0207(19990910)46:1<45::aid-nme662>3.0.co;2-k}.
\newblock URL
  \url{https://doi.org/10.1002/(sici)1097-0207(19990910)46:1<45::aid-nme662>3.0.co;2-k}.

\bibitem[J\"{u}ttler(1994)]{Jttler1994}
Bert J\"{u}ttler.
\newblock Visualization of moving objects using dual quaternion curves.
\newblock \emph{Computers {\&} Graphics}, 18\penalty0 (3):\penalty0 315--326,
  May 1994.
\newblock \doi{10.1016/0097-8493(94)90033-7}.
\newblock URL \url{https://doi.org/10.1016/0097-8493(94)90033-7}.

\bibitem[Park and Ravani(1997)]{Park1997}
F.~C. Park and Bahram Ravani.
\newblock Smooth invariant interpolation of rotations.
\newblock \emph{{ACM} Transactions on Graphics}, 16\penalty0 (3):\penalty0
  277--295, July 1997.
\newblock \doi{10.1145/256157.256160}.
\newblock URL \url{https://doi.org/10.1145/256157.256160}.

\bibitem[Haarbach et~al.(2018)Haarbach, Birdal, and Ilic]{Haarbach2018}
Adrian Haarbach, Tolga Birdal, and Slobodan Ilic.
\newblock Survey of higher order rigid body motion interpolation methods for
  keyframe animation and continuous-time trajectory estimation.
\newblock In \emph{2018 International Conference on 3D Vision (3DV)}. {IEEE},
  September 2018.
\newblock \doi{10.1109/3dv.2018.00051}.
\newblock URL \url{https://doi.org/10.1109/3dv.2018.00051}.

\bibitem[Dam et~al.(2000)Dam, Koch, and Lillholm]{Dam2000QuaternionsIA}
Erik~Bj{\o}rnager Dam, Martin Koch, and Martin Lillholm.
\newblock Quaternions, interpolation and animation.
\newblock 2000.

\bibitem[Svejda and Cechura(2015)]{Svejda2015}
Martin Svejda and Tomas Cechura.
\newblock Interpolation method for robot trajectory planning.
\newblock In \emph{2015 20th International Conference on Process Control
  ({PC})}. {IEEE}, June 2015.
\newblock \doi{10.1109/pc.2015.7169997}.
\newblock URL \url{https://doi.org/10.1109/pc.2015.7169997}.

\bibitem[Günaşti(2012)]{Gnati2012QuaternionsAT}
Gökmen Günaşti.
\newblock Quaternions algebra, their applications in rotations and beyond
  quaternions.
\newblock 2012.

\bibitem[Shoemake(1985)]{Shoemake1985}
Ken Shoemake.
\newblock Animating rotation with quaternion curves.
\newblock In \emph{Proceedings of the 12th annual conference on Computer
  graphics and interactive techniques - {SIGGRAPH} {\textquotesingle}85}. {ACM}
  Press, 1985.
\newblock \doi{10.1145/325334.325242}.
\newblock URL \url{https://doi.org/10.1145/325334.325242}.

\bibitem[Kuramoto(1975)]{Kuramoto}
Yoshiki Kuramoto.
\newblock Self-entrainment of a population of coupled non-linear oscillators.
\newblock In \emph{International Symposium on Mathematical Problems in
  Theoretical Physics}, pages 420--422. Springer-Verlag, 1975.
\newblock \doi{10.1007/bfb0013365}.
\newblock URL \url{https://doi.org/10.1007/bfb0013365}.

\bibitem[Lohe(2009)]{Lohe2009}
M~A Lohe.
\newblock Non-abelian kuramoto models and synchronization.
\newblock \emph{Journal of Physics A: Mathematical and Theoretical},
  42\penalty0 (39):\penalty0 395101, September 2009.
\newblock \doi{10.1088/1751-8113/42/39/395101}.
\newblock URL \url{https://doi.org/10.1088/1751-8113/42/39/395101}.

\bibitem[Ja{\'{c}}imovi{\'{c}} and Crnki{\'{c}}(2018)]{Jaimovi2018}
Vladimir Ja{\'{c}}imovi{\'{c}} and Aladin Crnki{\'{c}}.
\newblock Low-dimensional dynamics in non-abelian kuramoto model on the
  3-sphere.
\newblock \emph{Chaos: An Interdisciplinary Journal of Nonlinear Science},
  28\penalty0 (8):\penalty0 083105, August 2018.
\newblock \doi{10.1063/1.5029485}.
\newblock URL \url{https://doi.org/10.1063/1.5029485}.

\bibitem[Crnki{\'{c}} et~al.(2020{\natexlab{a}})Crnki{\'{c}},
  Ja{\'{c}}imovi{\'{c}}, Ja{\'{c}}imovi{\'{c}}, and
  Mijajlovi{\'{c}}]{Crnki2020}
Aladin Crnki{\'{c}}, Milojica Ja{\'{c}}imovi{\'{c}}, Vladimir
  Ja{\'{c}}imovi{\'{c}}, and Nevena Mijajlovi{\'{c}}.
\newblock Consensus and coordination on groups {SO}(3) and s3 over constant and
  state-dependent communication graphs.
\newblock \emph{Automatika}, 62\penalty0 (1):\penalty0 76--83, December
  2020{\natexlab{a}}.
\newblock \doi{10.1080/00051144.2020.1863544}.
\newblock URL \url{https://doi.org/10.1080/00051144.2020.1863544}.

\bibitem[Crnki{\'{c}} and Ja{\'{c}}imovi{\'{c}}(2018)]{Crnki2018}
Aladin Crnki{\'{c}} and Vladimir Ja{\'{c}}imovi{\'{c}}.
\newblock Data clustering based on quantum synchronization.
\newblock \emph{Natural Computing}, 18\penalty0 (4):\penalty0 907--911,
  November 2018.
\newblock \doi{10.1007/s11047-018-9720-z}.
\newblock URL \url{https://doi.org/10.1007/s11047-018-9720-z}.

\bibitem[Crnki{\'{c}} et~al.(2020{\natexlab{b}})Crnki{\'{c}}, Ivanovi{\'{c}},
  Ja{\'{c}}imovi{\'{c}}, and Mijajlovi{\'{c}}]{Crnki2020Swarms}
Aladin Crnki{\'{c}}, Igor Ivanovi{\'{c}}, Vladimir Ja{\'{c}}imovi{\'{c}}, and
  Nevena Mijajlovi{\'{c}}.
\newblock Swarms on the 3-sphere for online clustering of multivariate time
  series and data streams.
\newblock \emph{Future Generation Computer Systems}, 112:\penalty0 11--17,
  November 2020{\natexlab{b}}.
\newblock \doi{10.1016/j.future.2020.05.018}.
\newblock URL \url{https://doi.org/10.1016/j.future.2020.05.018}.

\bibitem[Ha and Park(2020)]{ha2020dynamical}
Seung-Yeal Ha and Hansol Park.
\newblock A dynamical systems approach for the shape matching of polytopes
  along rigid-body motions, 2020.

\bibitem[Hopf(1931)]{Hopf1931}
H.~Hopf.
\newblock Über die abbildungen der dreidimensionalen sphäre auf die
  kugelfläche.
\newblock \emph{Mathematische Annalen}, 104:\penalty0 637--665, 1931.
\newblock URL \url{http://eudml.org/doc/159489}.

\bibitem[Lyons(2003)]{Lyons2003}
David~W. Lyons.
\newblock An elementary introduction to the hopf fibration.
\newblock \emph{Mathematics Magazine}, 76\penalty0 (2):\penalty0 87--98, April
  2003.
\newblock \doi{10.1080/0025570x.2003.11953158}.
\newblock URL \url{https://doi.org/10.1080/0025570x.2003.11953158}.

\bibitem[Sangwine and Bihan(2008)]{Sangwine2008}
Stephen~J. Sangwine and Nicolas~Le Bihan.
\newblock Quaternion polar representation with a complex modulus and complex
  argument inspired by the cayley-dickson form.
\newblock \emph{Advances in Applied Clifford Algebras}, 20\penalty0
  (1):\penalty0 111--120, August 2008.
\newblock \doi{10.1007/s00006-008-0128-1}.
\newblock URL \url{https://doi.org/10.1007/s00006-008-0128-1}.

\bibitem[Estefo et~al.(2019)Estefo, Simmonds, Robbes, and Fabry]{Estefo2019}
Pablo Estefo, Jocelyn Simmonds, Romain Robbes, and Johan Fabry.
\newblock The robot operating system: Package reuse and community dynamics.
\newblock \emph{Journal of Systems and Software}, 151:\penalty0 226--242, May
  2019.
\newblock \doi{10.1016/j.jss.2019.02.024}.
\newblock URL \url{https://doi.org/10.1016/j.jss.2019.02.024}.

\end{thebibliography}





\end{document}